\documentclass[sigconf]{acmart}

\setcopyright{none}
\renewcommand\footnotetextcopyrightpermission[1]{}

\usepackage{fancyhdr}
\fancypagestyle{plain}{%
   \fancyhf{} %
   \fancyfoot[L]{ACM SIGCOMM Computer Communication Review}%
   \fancyfoot[R]{Volume 48 Issue 1, January 2018}%
}
\fancypagestyle{firstpagestyle}{%
   \fancyhf{} %
   \fancyfoot[L]{ACM SIGCOMM Computer Communication Review}%
   \fancyfoot[R]{Volume 48 Issue 1, January 2018}%
}

\usepackage[utf8]{inputenc}
\usepackage[T1]{fontenc}
\usepackage{lmodern}

\usepackage[english]{babel}
\usepackage{ae}
\usepackage{microtype}
\usepackage{url}
\usepackage{graphicx}
\usepackage{xcolor}
\usepackage[absolute,showboxes]{textpos}
\usepackage{enumitem}
\usepackage{framed}
\usepackage{textcomp}
\usepackage{subfigure}
\usepackage{balance}

\usepackage{pifont}

\usepackage{xspace}
\newcommand{\eg}{\textit{e.g.,}~}
\newcommand{\ie}{\textit{i.e.,}~}

\newcommand{\one}{({\em i})\xspace}
\newcommand{\two}{({\em ii})\xspace}

\makeatletter
\renewcommand{\paragraph}[1]{\vspace*{0.03in}\noindent{\bf #1.}\hspace{0.25ex \@plus1ex \@minus.2ex}}
\makeatother

\newcommand{\peering}{\textsf{PEERING}\xspace}

\begin{document}

\setlength{\TPHorizModule}{\paperwidth}
\setlength{\TPVertModule}{\paperheight}
\TPMargin{5pt}
\begin{textblock}{0.8}(0.1,0.02)    
     \noindent
     \footnotesize
     If you cite this paper, please use the CCR reference:
     A. Reuter, R. Bush, I. Cunha, E. Katz-Bassett, T.~C. Schmidt, M. W\"ahlisch.~2018. Towards a Rigorous Methodology for Measuring Adoption of RPKI Route Validation and Filtering. \emph{ACM SIGCOMM Computer Communications Review (CCR)} 48(1) (January 2018), pp. 19--27.
\end{textblock}

\title{Towards a Rigorous Methodology for Measuring Adoption of RPKI Route Validation and Filtering}

\author{Andreas Reuter}
	\affiliation{
	  \institution{Freie Universit\"at Berlin}
	}
	\email{andreas.reuter@fu-berlin.de}
\author{Randy Bush}
	\affiliation{
	  \institution{IIJ Research Lab / Dragon Research}
	}
	\email{randy@psg.com}
\author{Italo Cunha}
	\affiliation{
	\institution{\mbox{Universidade Federal de Minas Gerais}}
	}
	\email{cunha@dcc.ufmg.br}
\author{Ethan Katz-Bassett}
	\affiliation{
	  \institution{Columbia University}
	}
	\email{ethan@ee.columbia.edu}
\author{Thomas C. Schmidt}
	\affiliation{
	  \institution{HAW Hamburg}
	}
	\email{t.schmidt@haw-hamburg.de}
\author{Matthias W\"ahlisch}
	\affiliation{
	  \institution{Freie Universit\"at Berlin}
	}
	\email{m.waehlisch@fu-berlin.de}



\begin{abstract}
A proposal to improve routing security---Route Origin Authorization (ROA)---has
been standardized. A ROA specifies which network is allowed to announce a set
of Internet destinations. While some networks now specify ROAs, little is known
about whether other networks check routes they receive against these ROAs, a
process known as Route Origin Validation (ROV). Which networks blindly accept
invalid routes? Which reject them outright? Which de-preference them if
alternatives exist?

Recent analysis attempts to use uncontrolled experiments to characterize ROV adoption by comparing valid routes and invalid routes~\cite{gchss-trds-17}.
However, we argue that gaining 
a solid understanding of ROV adoption is impossible using currently available
data sets and techniques. 
Instead, we devise a verifiable methodology of controlled experiments for measuring ROV.  
Our measurements suggest that, although some ISPs are not observed using invalid routes in uncontrolled experiments, they are actually using different routes
for (non-security) traffic engineering purposes, without performing ROV. 
We conclude with presenting three AS that do implement ROV as confirmed by the operators.

\end{abstract}

\maketitle

\section{Introduction}
\label{sec:intro}

The Border Gateway Protocol (BGP)~\cite{RFC-4271} is responsible for establishing Internet routes, yet it does not check that routes are valid.
An autonomous system (AS) can hijack destinations it does not control by announcing invalid routes to them, either intentionally or unintentionally, as in the well-known accidental announcement of YouTube's address space by Pakistan Telecom~\cite{youtube-outage}.

Because this critical aspect of the Internet is vulnerable, there are
proposals to improve routing security~\cite{g-tlsir-14}, and one---the RPKI---is
standardized and is in early adoption.
The Resource Public Key Infrastructure (RPKI)~\cite{RFC-6480} is a specialized PKI to help secure Internet interdomain routing by providing attestation objects for Internet resource holders (\ie IP prefixes and AS numbers).
The RPKI publishes Route Origin Authorization (ROA) objects, each specifying which AS is allowed to announce an IP prefix.
Using ROA data, a BGP router can perform RPKI-based origin validation (ROV) verifying whether the AS originating an IP prefix announcement in BGP is authorized to do so \cite{RFC-6811} and labeling the route as valid or invalid. The validity of a route can be used as part of the router's local BGP policy decisions, \eg filtering routes that reflect invalid announcements or preferring valid ones.
While the RPKI is fairly populated with ROAs and growing~\cite{wms-tdbrh-12b,n-nrdm-15,ipb-mbror-15,wssmu-rtsrd-15}, adoption of ROV and filtering has been negligible, according to operator~gossip. 
A major reason for this is the lack of economic incentives. 
Since a significant share of invalid routes are due to misconfiguration~\cite{wms-tdbrh-12b}, adopting ROV and filtering can even have adverse effects such
as a loss of connectivity to legitimate network destinations.

A recent paper examined RPKI and ROV adoption from multiple angles, focusing on
the slow state of ROV adoption, the security implications of partial adoption,
and reasons for slow adoption. The paper also identifies an attack 
vector that exploits \emph{loose} ROAs to hijack traffic of a RPKI-secured prefix~\cite{gchss-trds-17}.
To capture the current state of limited adoption, the paper 
included a measurement study that claimed that most large AS had not deployed ROV, but that 9 of the 100~largest AS had. 
This result was based on observations of existing BGP routes from BGP route collectors, meaning that the experiments were uncontrolled. At a basic level, the approach finds an AS that originates both valid and invalid announcements, then identifies other AS that appear on paths towards the valid prefix but not on paths towards the invalid prefix. It then assumes these AS are performing ROV to filter invalid routes.

In this paper, we contribute a verifiable methodology  for measuring ROV after demonstrating that the above approach to identify ROV adoption, based on passive observation of routes in uncontrolled experiments~\cite{gchss-trds-17}, has three major limitations.
First, our measurements show that its  characterizations of
some networks change depending on which set of BGP collectors is used,
inferring ROV adoption in some cases when it definitely has not been deployed and not inferring it in some cases when it may have been deployed.
Second, the approach relies on invalid routes that happen to be announced, and so its coverage is limited by their rare nature~\cite{heilman-consent-of-the-routed-14}.
Third, we conducted supplemental measurements suggesting that most networks flagged by the approach (and by~\cite{gchss-trds-17}) as using ROV are actually avoiding invalid routes for unrelated (non-security) traffic engineering purposes, without checking ROV status. This means that adoption is likely even lower than suggested by the earlier study.
In fact, with only uncontrolled measurements of existing routes---the status quo for Internet research---it is impossible to differentiate between multiple feasible explanations.

To overcome challenges of measuring route origin validation~(\S~\ref{sec:challenges}) and the limitations of uncontrolled experiments~(\S~\ref{sec:uncontrolled}), we propose a method to accurately infer ROV policies using controlled experiments (\S~\ref{sec:controlled}) that manipulate both BGP announcements and the ROAs that apply to them. We discuss
the operational concerns related to ROV deployment and sketch a roadmap for further, more general
measurements~(\S~\ref{sec:controlled_roadmap}).
We provide initial results using our method, verified by ground truth.

Although ROV adoption is low and slow, our proposed method allows accurate, longitudinal observation of ROV adoption across the Internet.

\vspace{-0.25cm}
\section{The Challenges of Measuring \\ Route Origin Validation}
\label{sec:challenges}

\paragraph{Limited visibility} Measuring the deployment of ROV is challenging 
because of very limited visibility of routing decisions, which has multiple causes. First,  
an AS does not propagate every path it knows, instead selecting a best path to
each destination prefix and then choosing for each neighbor whether to export
that best path. So, BGP hides information by only forwarding a subset of
available paths to a subset of neighbors. Second, an AS can use arbitrary policy to select a best path
and to decide which neighbors to forward it to, and this policy is opaque. The
policy may reflect concerns such as business relationships and traffic engineering,
as well as route origin validity, and so it can be very difficult to discern
the cause of any observed decision. Third, the interactions of these policies
can influence the decisions of seemingly uninvolved AS, meaning that
it is not enough to observe a path before and after a change to understand which
AS caused the change~\cite{jccka-pirci-13}. Fourth, as researchers, 
we typically have a limited view of the Internet, with projects such as RIPE RIS~\cite{riperis} and RouteViews~\cite{routeviews}  collecting routes from a small number of AS, many of which only provide their routes to a limited set of destinations~\cite{opwzz-segti-08}. This makes it hard to locate where routes diverge or whether differences are due to actual filtering or simply lack of visibility.

\paragraph{Lack of controlled experiments} We distinguish between two experimental methods, controlled and uncontrolled. In a controlled experiment, researchers vary one factor of interest (whether a route is valid) while fixing other factors, then measure the outcome (which route an AS uses), observing how networks route under different scenarios of interest to the current research question~\cite{jccka-pirci-13, szcfk-pau-14}.
In an uncontrolled experiment, the factor of interest varies outside the control of the researchers and independent of the current research question (AS on the Internet happen to announce a mix of valid and invalid routes), and researchers measure outcomes. 

Our classification of controlled versus uncontrolled describes experiments (\emph{how to test a hypothesis}).  It is orthogonal to the classification of passive versus active measurements (\emph{how data are collected}), and passive versus active measurements are orthogonal to control plane versus data plane measurements (\emph{what data are collected}). 
With uncontrolled experiments, inferring root causes of routing decisions is challenging because pinpointing the reason for the decision (\eg RPKI policy or traffic engineering) is difficult when path attributes and RPKI data cannot be manipulated independently to observe their impact on decisions.

\paragraph{Implementation variations}
Uncontrolled experiments are most challenged when the baseline of the system is unclear or complex.
The deployment of ROV introduces additional variations in implementation and configuration (\eg ROA propagation delay~\cite{mppba-tfebs-12}, route revalidation because of ROA change) that have not yet been explored but likely affect measurement outcome.

\section{Revisiting Uncontrolled experiments}
\label{sec:uncontrolled}

A previous approach for detecting ROV deployment used uncontrolled passive measurements \cite{gchss-trds-17}.
This study did not release the code and data sources needed to reproduce it.
In this section, we try to replicate some of the results and analyze how reliably the method leads to the conclusions.
We show that the limited view provided by vantage points can lead to incorrect identification of ROV non-adoption and ROV adoption.
Our analysis shows that differences in AS paths towards valid and invalid announcements are mainly a measurement artifact, instead of evidence for filtering.

To be clear on terminology, a \emph{route collector} is a BGP router that peers with border routers of various AS, each of which we refer to as a \emph{vantage~point}~\cite{rwmpb-lymmi-11}.

\subsection{Uncontrolled Method}
\label{sec:passive-method}
The previous approach uses available BGP dumps and RPKI data to estimate a lower bound for ROV non-adoption and identify ROV filtering~\cite{gchss-trds-17}. 
It compares AS paths taken by known ROV valid and known ROV invalid announcements from a single AS to a single vantage point.  If the paths differ, it assumes that the invalid announcement was filtered by ROV on the path taken by the valid announcement, causing the divergence.
This approach does not distinguish between a single router or an entire AS using ROV-based filtering, since it makes inferences based on the AS that appear on AS paths to vantage points.
The method analyzes routes exported by vantage points as follows:

\paragraph{Exclude AS observed to use invalid routes}
First, any AS that is found on a path of an invalid announcement is flagged as \emph{non ROV enforcing}. 
This assumes that any AS that accepts any invalid route accepts all invalid routes; \ie AS do not implement selective filtering or use other policies that can accept some invalid routes while filtering others.
An exception is made for invalid announcements originated by the vantage point's AS or by one of its customers~\cite{gchss-trds-17}, as an AS may make exceptions for its customers.

\paragraph{Identify AS that may be performing ROV filtering}
For each vantage point, the approach identifies all AS observed to originate at least one non-invalid (either valid or not in the RPKI database) prefix announcement and at least one invalid announcement.
It then compares each non-invalid path (from the origin to the vantage point) to each invalid path.
If there is exactly one AS that \one appears on the non-invalid path but not the invalid path, and \two has not been flagged as \emph{non ROV enforcing}, the approach marks it as an \emph{ROV candidate} for announcements from that origin.

For example, the vantage point $V$ might observe the following paths for the non-invalid prefix announcements $P_{1-2}$ and invalid prefix announcements $P_{3-4}$ advertised from origin~$O$:

\medskip
\begin{tabular}{rr}
\hspace{1cm} $P_{1}: O\to A\to C\to V$ & \hspace{1.1cm} \texttt{not found} \\
$P_{2}: O\to A\to E\to V$ & \texttt{valid} \\
$P_{3}: O\to A\to D\to V$ & \texttt{invalid} \\
$P_{4}: O\to A\to D\to V$ & \texttt{invalid}\\
\end{tabular}
\medskip

\noindent In this case, AS $C$ and AS $E$ are marked as \emph{ROV candidates} for origin $O$, unless they have been previously marked as \emph{non ROV enforcing}.

\paragraph{Select filtering AS}
The approach then counts the number of origins for which it marked an AS as an \emph{ROV candidate} and, 
following previous work \cite{gchss-trds-17}, classifies an AS marked for at least 3 origins as \emph{ROV enforcing}.

\subsection{Data Set and Comparison with Current Findings}
\paragraph{Data Set}
The previous study~\cite{gchss-trds-17} specifies the data set they have used to be from July 2016, collected from 44 Routeviews vantage points. It does not mention which vantage points
explicitly.
Our analysis is based on BGP RIB dumps gathered from all route collectors of the RIPE RIS and Routeviews projects from October 25th 2016, 16:00 UTC. This data set includes 27GB of exported
routes from 960 vantage points, a larger data set than the previous study has used. 
The data exported by the vantage points includes routes to both IPv4 and IPv6 prefixes.

\paragraph{Reproducing existing methodology}
We reproduced the methodology from the description in the previous study~\cite{gchss-trds-17}, since the original code is not available.
Analyzing our complete data set using the uncontrolled method, it classifies the following AS as \emph{ROV enforcing}:
\begin{center}
$AS8100$
$AS25761$
$AS17819$
$AS262150$
\end{center}
None of these AS is among the top-100 AS, based on the CAIDA AS rank~\cite{caida-asrank}.
This result differs from previous measurements \cite{gchss-trds-17}, which used a different set of RIB dumps to conclude that 9 of the top 100 AS enforce ROV.
We want to better understand the validity of the method and why results vary significantly.

\subsection{Impact of Limited Vantage Point Sets}
\label{sec:impact-vp-sets}
When we run the same analysis on a subset of our data, such as data from a single route collector, the results differ.
For example, the \texttt{routeviews-equix} collector has a feed from 34 vantage points, yet running the same analysis just on this feed results in zero AS marked as \emph{ROV enforcing}.
In contrast, the \texttt{routeviews-wide} collector has feeds from only 4 vantage points, but shows the following AS as \emph{ROV enforcing}:
\begin{center}
$AS48237$
$AS262150$
$AS3786$
\end{center}

\begin{figure}
\centering
	\includegraphics[angle=90,width=0.1\linewidth]{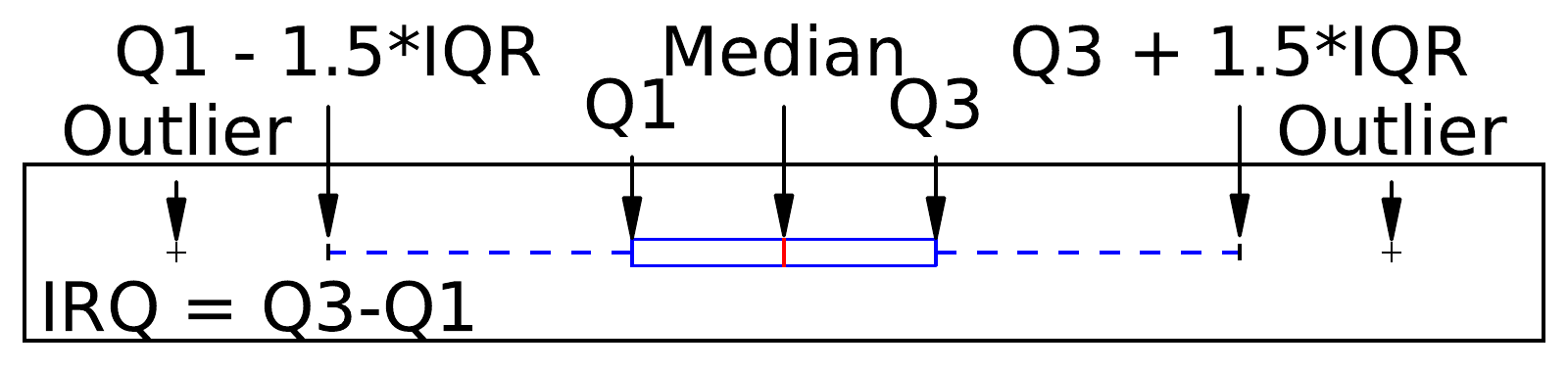}
  \subfigure[]{\includegraphics[width=0.25\linewidth]{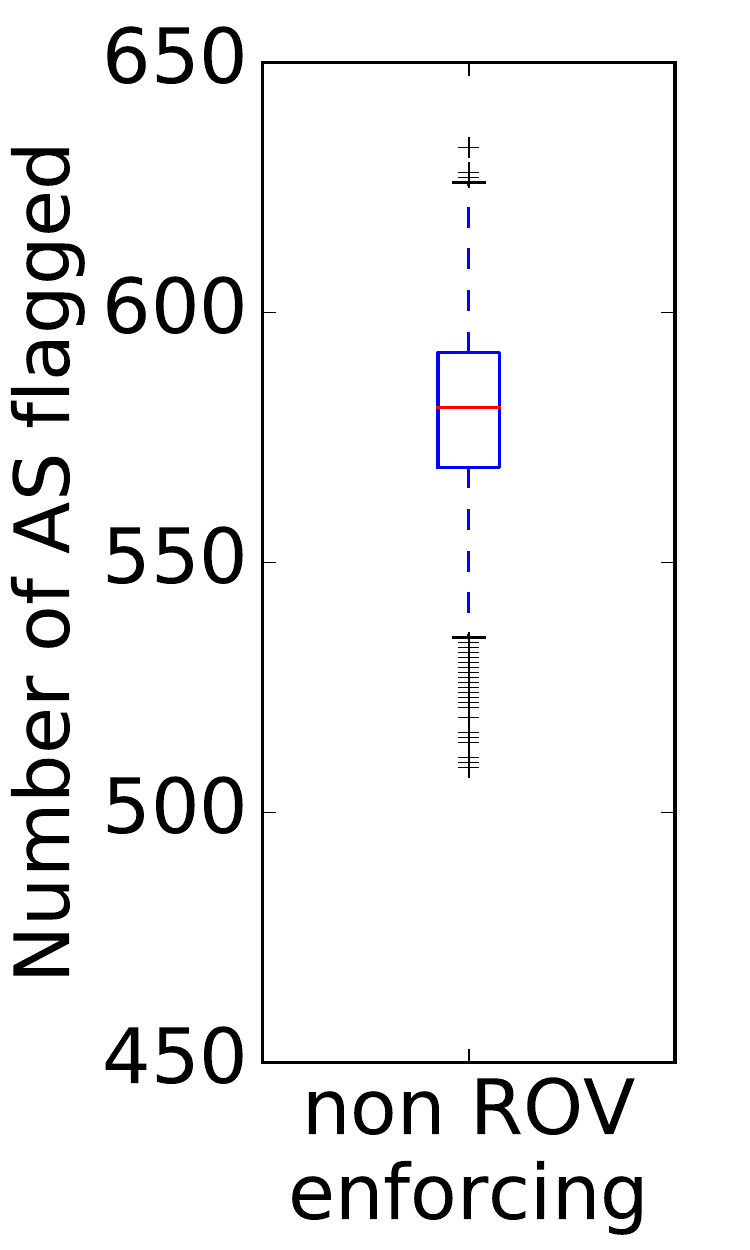}
    \label{fig:passive-rand-sampling-non-rov}}
  \subfigure[]{\includegraphics[width=0.25\linewidth]{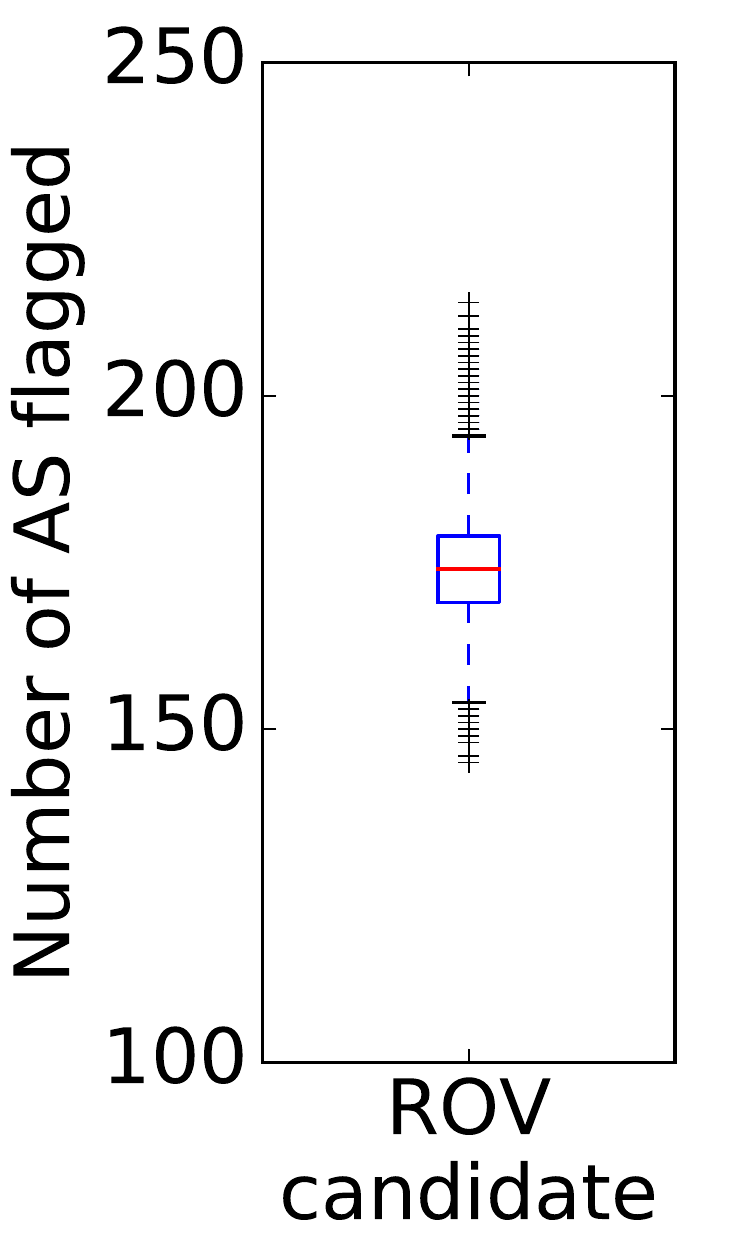}
    \label{fig:passive-rand-sampling-rov-cand}}
  \subfigure[]{\includegraphics[width=0.25\linewidth]{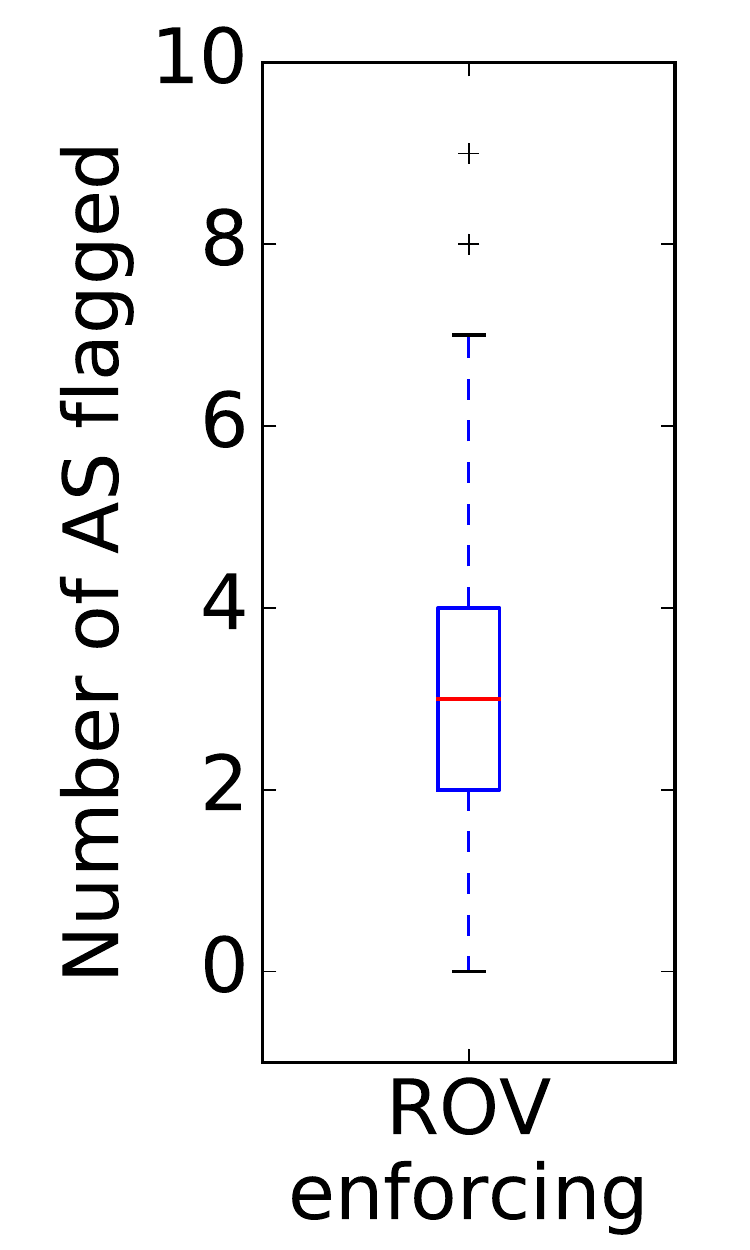}
    \label{fig:passive-rand-sampling-rov-enf}}
  \caption{Uncontrolled, passive measurements: Statistical impact of vantage points on the number of identified AS (5,000 samples of 44~randomly selected vantage points).}
  \label{fig:passive-rand-sampl}
\vspace{-0.4cm}
\end{figure}
Out of those 3 AS, $AS48237$ and $AS3786$ are both found on the AS paths of invalid routes when considering data from the \texttt{route-views4} collector, 
classifiying them as \emph{non ROV enforcing}, contradicting the previous \emph{ROV enforcing} classification.
This shows that using the uncontrolled methodology some AS might be (mis)classified as \emph{ROV enforcing} if the invalid announcements they propagate are not visible in the data
set, leading to false positives. We define an AS as a false positive if it is classified 
as \emph{ROV enforcing} using data set $d_{1}$, but classified as \emph{non ROV enforcing}
using data set $d_{2}$, with $d_{1}$ encompassing $d_{2}$.
On the other hand, some AS that are classified as \emph{ROV enforcing} in a more complete data set might not be visible enough in a smaller data set to result in 
a classification as \emph{ROV enforcing}, leading to false negatives compared to the full data set. For example, using the complete data set, the approach marks AS8100 as 
an \emph{ROV candidate} for origins AS6921, AS46562, and AS46261. It is thus flagged as \emph{ROV enforcing}. When looking only at the data from the \texttt{routeviews-wide} 
collector, AS8100 is only marked as \emph{ROV candidate} for a single origin, AS46562, and thus it is not classified as \emph{ROV enforcing}.

\begin{figure}
\centering
	\subfigure[AS flagged as ROV enforcing and number of false positives]{\includegraphics[width=0.22\textwidth]{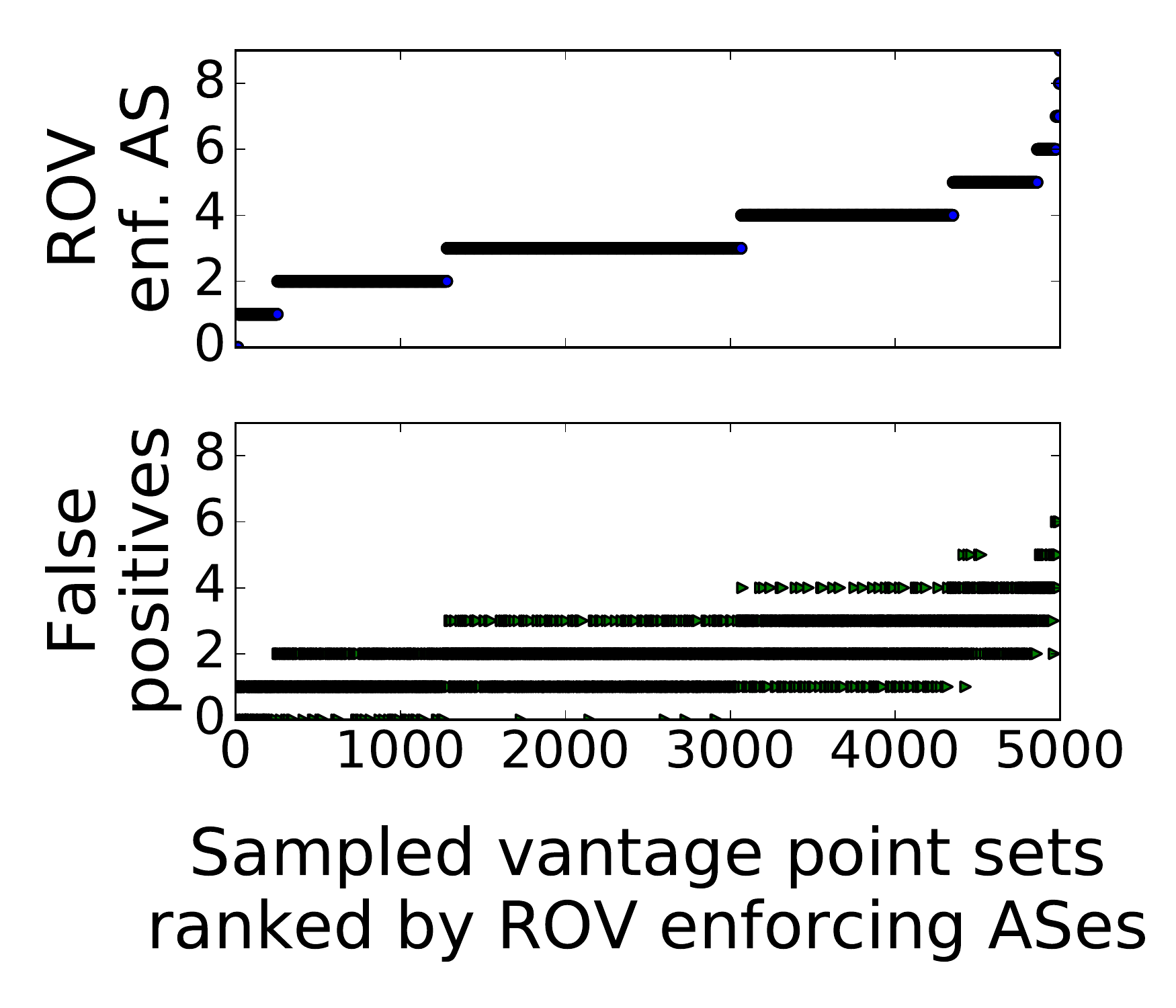}
	 \label{fig:rs-false-positives-scatter}}
	\subfigure[Relative frequency of false positives]{\includegraphics[width=0.22\textwidth]{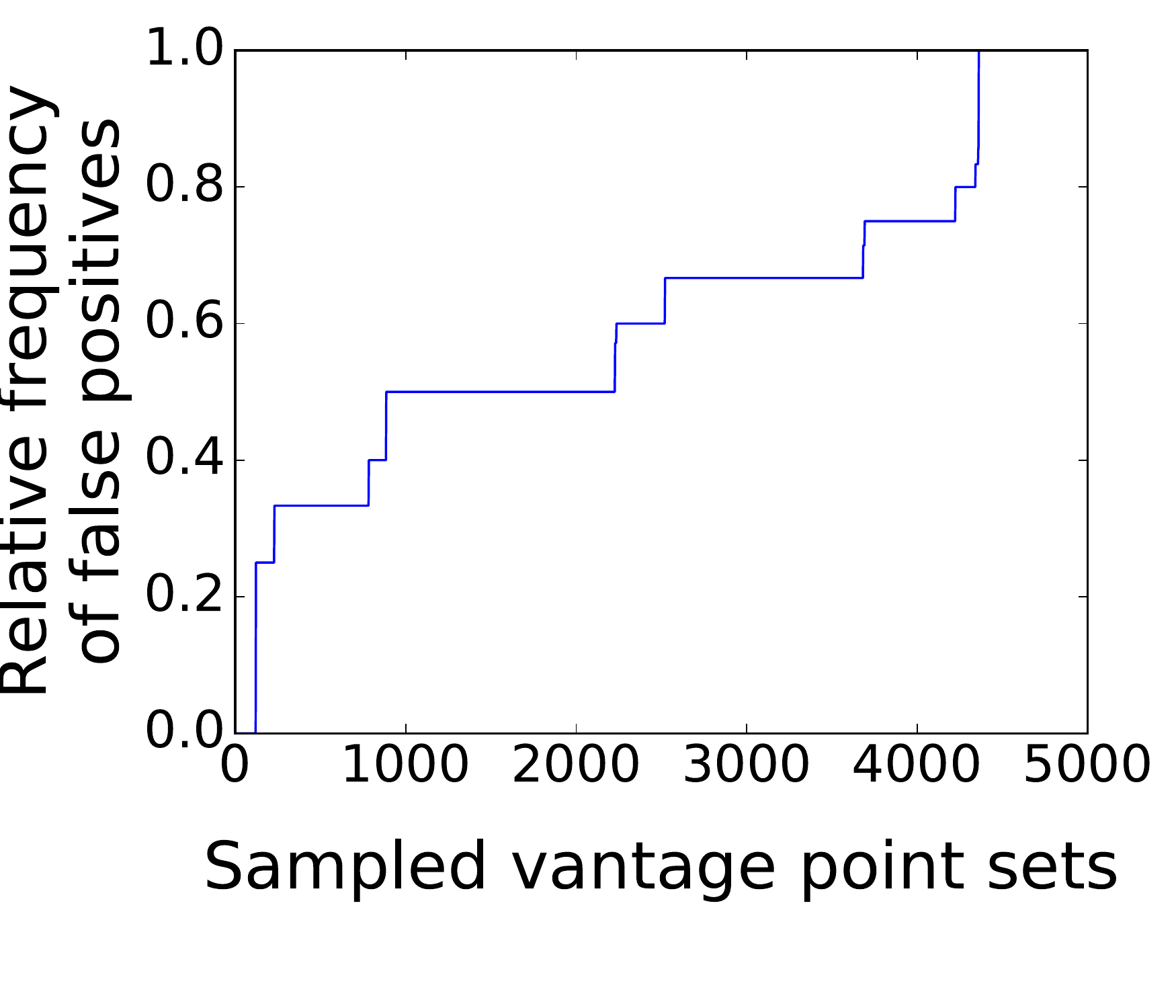}
	 \label{fig:rs-false-positives-fraction}}
	 \caption{Number of misclassified \emph{ROV enforcing} AS in different vantage point~sets.}
	 \label{fig:rs-false-positives}
\end{figure}

This shows that results vary significantly depending on which set of vantage points the data is taken from.
To quantify the impact of this we select 44~Routeviews vantage points (the number used in previous work~\cite{gchss-trds-17}) and calculate the number of AS identified in each step of the method (see \S~\ref{sec:passive-method}).
Figure~\ref{fig:passive-rand-sampl} summarizes statistical properties (quartiles, extreme non-outliers, and outliers) of 5,000 random samples of 44 vantage points,
showing that, even for a fixed number of vantage points, results can vary widely depending on which vantage points are used.
Results for a single selection of vantage points may not reliably determine a lower bound of either deployment or non-deployment.
Figure~\ref{fig:rs-false-positives} depicts the number of false positives of \emph{ROV enforcing} AS, those classified as enforcing given sampled subsets of 44~vantage points but non-enforcing based on the global data set.
For 82\% of the samples, the ratio of false positives is 50\% or more.

\paragraph{Conclusion}
Using BGP RIB dumps as a basis for uncontrolled measurements of ROV filtering (or non-filtering) is problematic.
It makes inferences based on routes visible in the selected dumps, but lacks complete visibility of the global Internet, leading to misclassification. 

\subsection{Impact of Limited Prefix Visibility at VPs}
\label{sec:impact-vp-prefixes}
Recall that the existing approach to identify ROV filtering compares paths for invalid announcements with paths for non-invalid (\ie valid or unknown) announcements.
We have shown that the selection of vantage points has a major impact on classification using this approach. 
As the approach uses pairs of non-invalid and invalid announcements, it relies on vantage points receiving such announcements from enough origins to reveal their policies.

\begin{figure}
\centering
  \subfigure[Invalid prefix announcements]{\includegraphics[width=0.48\linewidth]{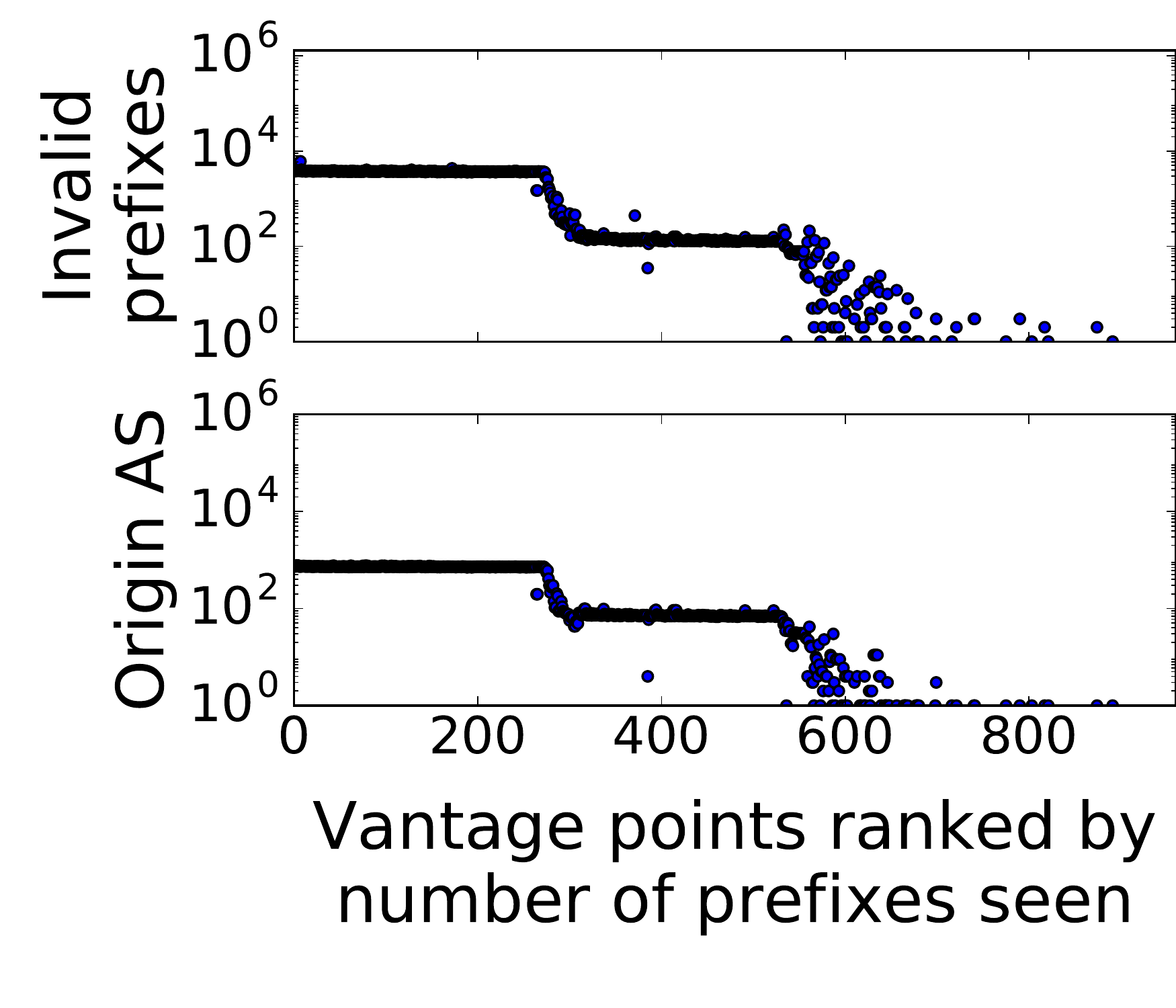}
    \label{fig:prefixvis-invalids}}
  \subfigure[Relative prefix completeness seen per vantage point]{\includegraphics[width=0.48\linewidth]{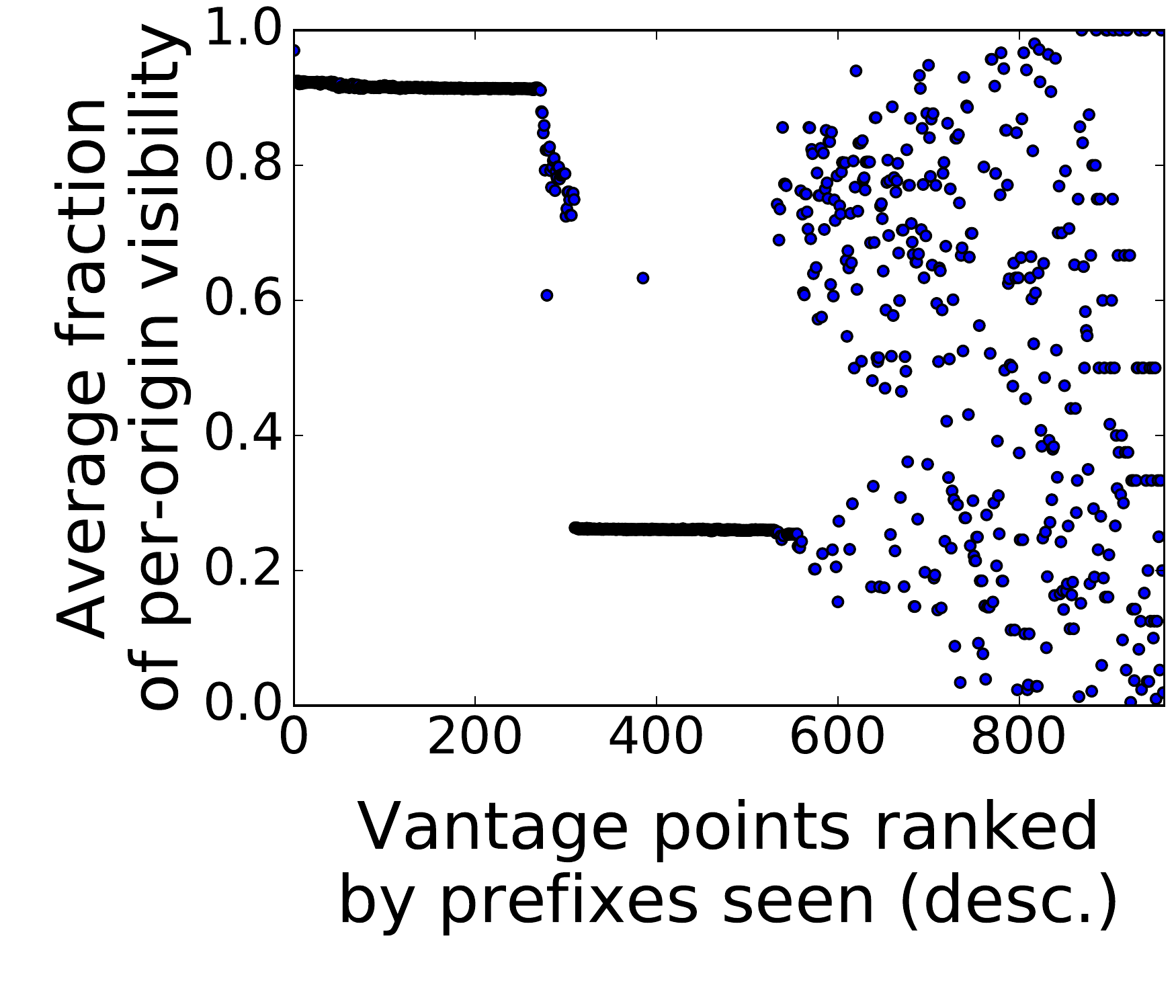}
    \label{fig:origin-coverage}}
    \caption{Number of prefixes and origin AS observed by RIPE and Routeviews.}
  \label{fig:prefixvis}
\end{figure}

Combining all dumps from the RIPE RIS and Routeviews projects, we have data from 960~vantage points. But, not all vantage points provide routes to the same set of prefixes. Some vantage points have a near global view, while some have routes for only a very limited number of prefixes.

For each vantage point, Figure~\ref{fig:prefixvis-invalids} shows the number of prefixes received via invalid announcements (top) and the number of distinct origins originating these announcements (bottom).
Though some vantage points provide routes for invalid prefix announcements to nearly 1000 distinct origin AS, more than 36\% of the vantage points see less than the needed 3~AS originating invalid prefix announcements. 
This observation is independent of the RPKI deployment state.
Figure~\ref{fig:origin-coverage} shows the relative ratio of visible prefixes per origin and vantage point.
Those vantage points that see many prefixes lack a complete view with respect to all prefixes per origin.

\paragraph{Conclusion}
Assuming one applies the method with only a subset of VPs as in the previous work \cite{gchss-trds-17}, selecting vantage points with very limited prefix visibility misses a significant portion of origin AS, and thus underestimates the set of \emph{ROV candidates} and can lead to misclassification.

\subsection{Impact of Limited Control}
\label{sec:uncontrolled-te}

Just because a vantage point uses different routes to reach a non-invalid and
an invalid prefix from the same origin does not imply that the difference is
caused by ROV-based filtering, as invalid and non-invalid advertisements might differ in attributes other than RPKI validity. We now investigate traffic engineering
as another possible explanation (unrelated to BGP security) for observed
differences.
For a multi-homed AS, a common technique to influence inbound traffic is to announce different prefixes to different upstreams.  These prefixes often overlap,
\eg an AS may announce a more specific prefix (a /24) via upstream~$A$ and the covering prefix (a /16) via upstream~$B$ to shift traffic to $A$.
Studies comparing current ROAs to announced prefixes have shown that the major cause for invalid BGP announcements is issuing a ROA only for a prefix and then announcing subprefixes \cite{wms-tdbrh-12b,ipb-mbror-15,draft-yossigi-rpkimaxlen-00} which are not covered by the ROA.
Announcing the /16 and /24 to separate providers then results in two routes, one valid and one invalid, diverging on the first hop of the AS path.

\begin{figure}
\centering
	{\includegraphics[width=0.90\linewidth]{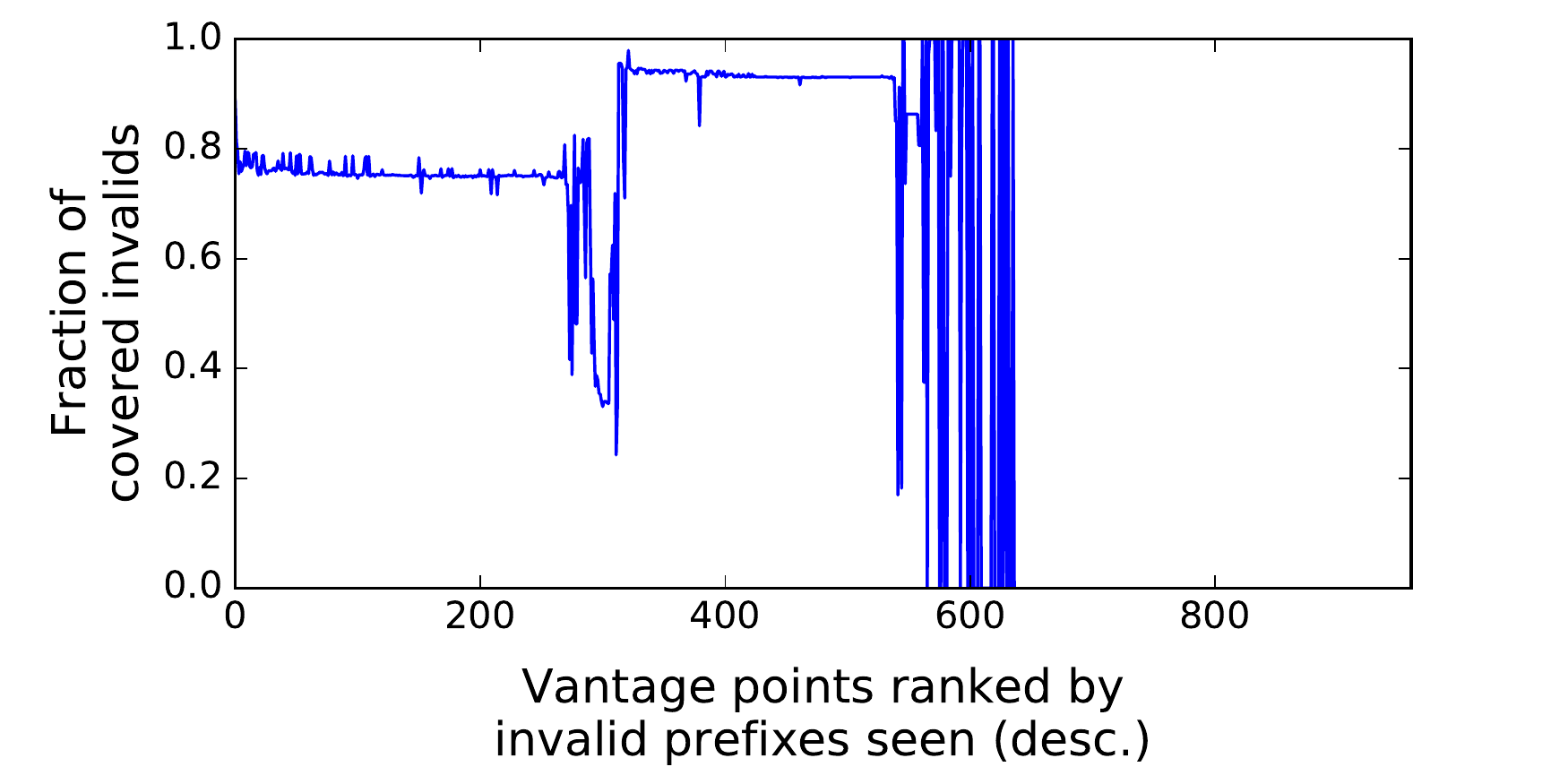}
	\caption{Fraction of invalid prefixes covered by a valid less-specific prefix from the same origin.}
	 \label{fig:invalids-len-coverage-local}}
\end{figure}

For each vantage point, Figure~\ref{fig:invalids-len-coverage-local} shows the fraction of prefixes from invalid announcements that are covered by a prefix from a non-invalid announcement from the same origin.
An invalid prefix only counts as \emph{covered} if the vantage point sees both the route to the invalid prefix and a route for the covering non-invalid prefix.
For the vantage points between $x=[0,~275]$, roughly 80\% of prefixes from invalid announcements are covered by a non-invalid from the same origin. This strongly suggests that the prefixes
are invalid because of incorrect ROA configuration and the announcements perhaps subject to traffic engineering.

\begin{figure}
\centering
        {\includegraphics[width=0.9\linewidth]{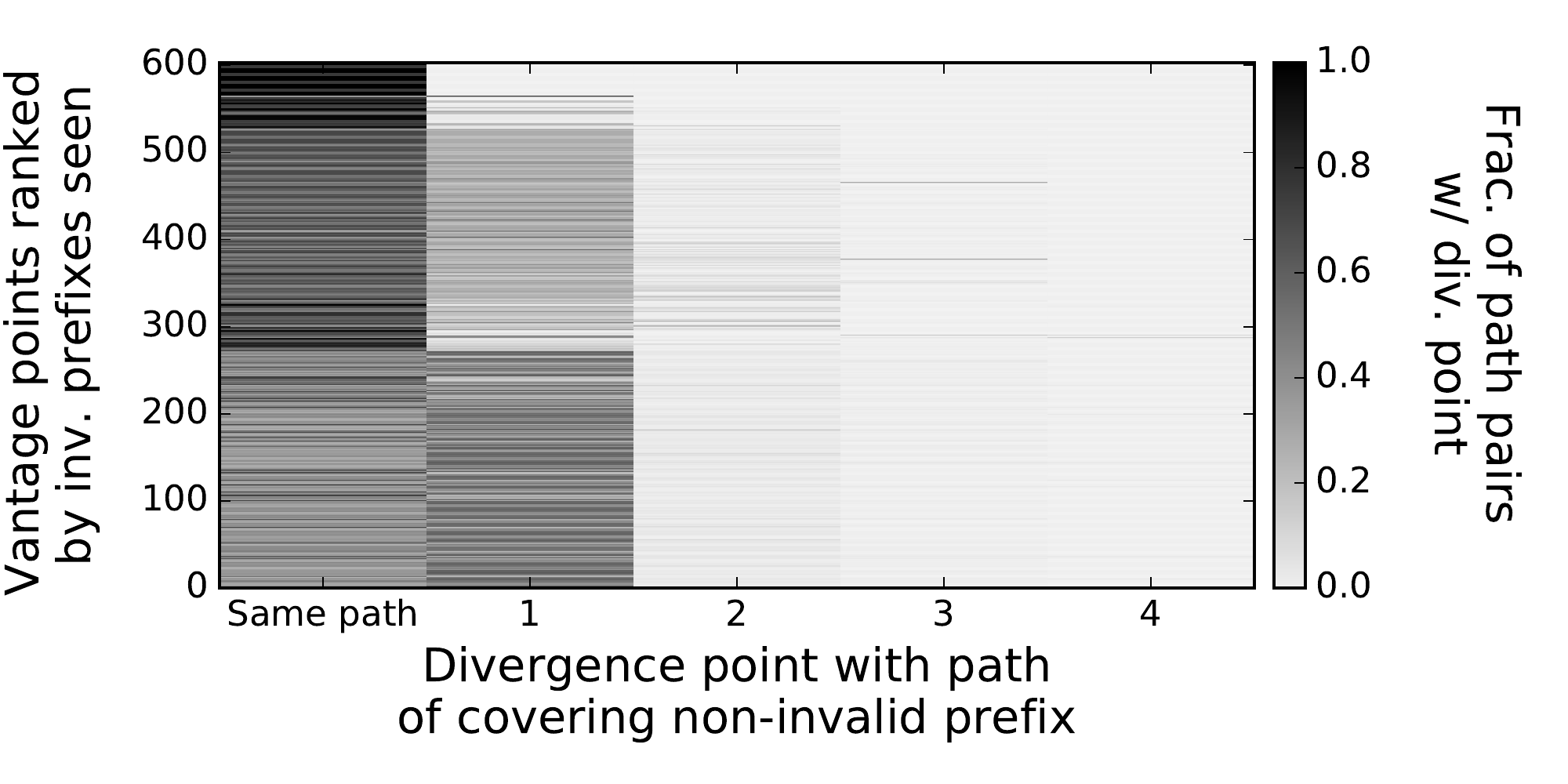}
  \caption{Divergent AS hop distribution of invalid prefixes with covering non-invalid prefixes of same origin.}
         \label{fig:invalid-len-divergence}}
\end{figure}

Next, we investigate where the vantage points' paths to these invalid prefixes diverge from their paths to the covering prefixes.
Figure~\ref{fig:invalid-len-divergence} shows the distribution of divergence points for all vantage points. The $y$-axis sorts the vantage points by the number of invalid prefixes they provide routes to.  The coloring of the $x$-axis depicts the fraction of these paths that diverged a given number of hops from the origin.
The majority of AS paths of invalid routes either share the AS path of the covering non-invalid ($x$=``Same path'') or diverge at the first hop, as would occur with traffic engineering.

\paragraph{Conclusion} 
ROV-based filtering is not the only plausible explanation for instances of vantage points using different routes to reach non-invalid and invalid prefixes from the same origin. We found that most instances display signatures of traffic engineering, and, during our study, we also observed a router selecting different routes from the same origin AS due to route age (a BGP tiebreaker).

\section{Controlled Experiments}
\label{sec:controlled}

With uncontrolled, passive experiments, it can be impossible to determine whether an AS is actually filtering  or whether it is not using 
invalid advertisements because of other attributes. 
Further, the AS where the divergence occurred need not be the one that made a different decision, as it could have been presented with different options for its decisions.
To overcome misclassification, experiments must clearly establish whether decisions stem from ROA status. 

Controlled experiments provide a means to establish this causation despite our limited visiblity into routing decisions. Based on the challenges in measuring ROV adoption~(\S\ref{sec:challenges}) and our experiences evaluating the existing approach~(\S\ref{sec:uncontrolled}), we arrive at the following requirements for a more reliable methodology.

\paragraph{Experiments must be long-lived} Adoption is likely to be slow,
  and  may be bursty, driven by various initiatives and technologies.
  Measurements must be rerun periodically.

\paragraph{Experiments must be active and controlled} Passive observations
    of existing announcements are insufficient to determine a policy since 
    we do not know precisely how the announcements are being made (\ie traffic
    engineering or not) and route collectors may not provide the right vantage 
    to locate filtering.
    Furthermore, we need to coordinate announcements with ROA changes to 
    precisely expose policies.

\paragraph{Experiments require rich BGP connectivity} From a single
  vantage point, it is difficult to infer which network along a path is
  filtering an announcement.

\subsection{Basic Approach}\label{approach}

We describe an approach based on 
active, controlled manipulation of BGP announcements and RPKI ROAs. 
We use the \peering testbed, which allows us to make BGP announcements for prefixes we control from \peering sites around the world to the hundreds of networks it peers with~\cite{szcfk-pau-14}.
We use multiple /24 prefixes from the same /16 block.
These prefixes share the same route object in the Internet Routing Registry.
To control ROAs, we run a grandchild RPKI Certificate Authority (CA) in the RIPE region, enabling us
to programmatically issue and revoke Resource Certificates and ROAs.
Common RPKI cache server implementations by default update at an interval of
10 to 60 minutes.
To guard against uncommonly long ROA propagation delays, we conservatively keep every configuration (set of BGP announcements and ROA states) in place for eight hours.

In our basic approach, an AS must fulfill two assumptions to allow us to unambiguously determine whether the AS is using ROV-based filtering:
  \one \emph{connected-assumption.} The network peers with \peering, either directly or using a route server.
  \two\emph{visibility-assumption.} The network offers some means to check the BGP route it uses to reach an Internet destination, either via a Looking Glass or via 
  a vantage point.
While the connected assumption is limiting, it is necessary to maintain accuracy, relaxing it to allow networks that are not peers of \peering introduces 
ambiguity. We discuss the possibility of relaxing the connected assumption, as well as the visibility assumption, in section~\ref{sec:controlled_roadmap}.

We announce two prefixes via \peering (AS47065), a \emph{reference prefix} $P_R$ and an \emph{experiment prefix}~$P_E$.
We periodically change RPKI state for the experiment prefix, using an additional origin AS $O$ to alternate between the following configurations:
\begin{enumerate}
  \item[(C1)] ROA specifies AS47065 is \texttt{valid} for $P_R$ and $P_E$, so both announcements are valid.
  \item[(C2)] ROA specifies AS47065 is \texttt{valid} for $P_R$, AS~$O$ is
  \texttt{valid} for $P_E$. AS47065's announcement of $P_E$ is invalid.
\end{enumerate}
We check the routes a vantage point chooses to both prefixes during both configurations.
The reference prefix always has a \emph{valid} RPKI state so should not be filtered via ROV, and so we omit any vantage points at which $P_R$ is not visible.
We expect both prefixes to be treated the same as long as both announcements are valid, and so we omit a vantage point if it uses different routes during configuration C1.
Analysing only data from vantage points that pass both these requirements eliminates the problem of \textbf{limited visibility}, since there is no missing data anymore.
We then check the routes a vantage point has chosen after the announcement of the expermiment prefix becomes \emph{invalid}. 
Three observations might occur:
\emph{(O1)} $V$ has the same route for both prefixes $P_E$ and~$P_R$.
\emph{(O2)} $V$ has a different route for prefix $P_E$.
\emph{(O3)} $V$ has no route to $P_E$.

In the cases of O2 and O3, we know that this route change \emph{must} be because of the RPKI status change. Had it been for another reason we would 
expect a change in route for the reference prefix as well. The reference prefix combined with the ROA changes thus all but eliminates the problem of 
\textbf{limited control}. The experiments are repeated continuously to confirm the behaviour is consistent.

\subsection{Measurements and Results}

\paragraph{Experiment Reach}
The experiments were conducted using \peering BGP routers in Amsterdam and Seattle. The device in Amsterdam peers with 589 different AS,
either directly or via a routeserver at AMS-IX. The device in Seattle peers with 179 different AS either directly or via a routeserver
at SIX. In total, via these two location \peering peers with 730 AS. Out of these 730 AS, only 138 AS peer with a RIPE RIS or Routeviews route collector.
Out of those 138 vantage points, 68 actually export direct routes for prefixes announced by \peering.

\paragraph{Results}
These experiments were performed February 20-27, May 11-17, and August 1-7, 2017.
In our experiments in February and May~2017, we found AS8283, AS50300, and AS59715 were using ROV to filter invalid announcements.
AS8283 and AS50300 comply with both of our assumptions. The experiments in August show AS50300 and AS59715 to be filtering, but not AS8283. 

AS8283 was identified based on observation (O3), and AS50300 based on (O2).
It is worth noting that AS50300 only filtered routes learned via a route server at the Amsterdam exchange (AMSIX). 
This contradicts one of the assumption in the methodology studied in section~\ref{sec:uncontrolled}, whereas it is assumed
that an AS found on the AS path of an invalid route does not use ROV based filtering.

AS59715 was not directly connected to \peering but lead to (O3).
For all three AS we contacted the operators via email and they confirmed that they used ROV based filtering. In the case 
of AS8283, they confirmed that they had shut off ROV based filtering for technical reasons in July 2017. This confirms our
findings from August 2017.

Relaxing the connected-assumption in the basic approach lead to ambiguity since multiple AS can be on the path between \peering and a vantage point.
To deal with cases such as AS59715 precisely, we propose a roadmap for a more general approach in the next~section.

\section{Roadmap}
\label{sec:controlled_roadmap}

\subsection{Operational Concerns}

\paragraph{ROA Propagation Time}
Analysis of our experiments has shown that the time for some AS to receive newly issued ROAs can be up to 8 hours or more.
We have also observed that the propagation time for some AS is inconsistent and varies by up to 2 hours. It is not clear yet whether this
is due to RPKI cache servers updating infrequently or due to BGP routers using excessively long refresh intervals.

\paragraph{Considering implementation variations} Active RPKI experiments require a careful check of router implementations~\cite{RFC-7128}.
For a router to perform ROV when an existing route changes from valid to invalid (due to an RPKI change), the BGP implementation must \one 
receive the new ROA payload and \two recalculate the best path for this existing entry. 
We verified that Cisco and Juniper implementations do recalculate the best path upon ROA changes, however there are corner cases where certain Cisco 
implementations do not re-apply route-maps that change BGP path attributes based on RPKI validation.
This might lead to a filtering AS to go unnoticed by our basic approach. To detect such cases we have set up a second set of experiments in which 
the BGP announcements are withdrawn prior to the ROA changes and then
reannounced once the new ROAs have propagated. We have timed these announcements 
in such a way to minimize risk of BGP dampening.
So far we have found no additional AS to be filtering in these experiments.  

Analyzing router implementations in more detail, analyzing the consistency among RPKI cache servers, and measuring ROA propagation time to routers is part of our ongoing~work.

\subsection{Extending Controlled Experiements}

Going forward, we plan to generalize our approach by conducting additional experiments, but also relaxing our assumptions without sacrificing the precise conclusions enabled by tightly controlled experiments.

\paragraph{Relaxing \texttt{connected-assumption}}
Suppose the target does not connect directly to \peering and has no
route to $P_E$. It might check
ROAs or might not receive a route from any neighbor.
To narrow our policy inferences, we use two techniques.
First, we iteratively target networks in a breadth-first
search outwards from a \peering site, similar to an approach that we used to uncover
(non-security-related) routing policies \cite{anwar15interdomain}.
Second, we will make multiple observations and only consider inferences consistent with all observations. 
We will make multiple observations both by using vantage points across the Internet and by targeting a network with different announcements.
We can vary the announcements by changing which \peering sites we use, which peers we announce to, and
what BGP attributes we use to influence route
selection and propagation. 

\paragraph{Relaxing \texttt{visibility-assumption}} Lacking a BGP
feed from a network, we can measure the data plane. This is
straightforward if it has a traceroute server or RIPE Atlas
probe~\cite{ripeatlas}. If not, we can ping a destination in
the target network and check the \peering site the reply arrives at, or
use our Reverse Traceroute~\cite{revtr}.

\paragraph{Inferring complex RPKI policies}
A network may prefer valid routes over invalid but not drop invalid routes.
In order to test for such policies, experiments must fulfil an additional requirement:

\paragraph{Experiments require competing announcements} To identify
  \texttt{prefer-valid} policies, we need multiple simultaneous
  announcements for the same addresses. Since a single BGP session
  generally allows only a single announcement, the experiment must
  include sessions with multiple peers.

In order to test for such policies we announce two prefixes $P_R$ and $P_E$ identically, 
each from two different locations with two different ASN (61575 and 61576).

Initially, we configure ROAs in such a way that announcements for $P_R$ from
both origin AS are valid, and announcements for $P_E$ are invalid.
We then announce $P_R$ and $P_E$ \emph{exclusively} from AS61575 and check whether vantage
points will receive these routes. We repeat the same step for AS61576. 
Once we have confirmed vantage points receive routes for $P_R$ and $P_E$ from
both origin AS, we announce both prefixes from both origin AS simultaneously.
Throughout the experiment, all announcements for the reference prefix $P_R$ will
stay valid, while announcements for $P_E$ will vary like this:

\begin{enumerate}
  \item[(C1)] ROA specifies AS61575. Announcement of $P_E$ from AS61575 is \texttt{valid}, from AS61576 is \texttt{invalid}.
  \item[(C2)] ROA specifies AS61576. Announcement of $P_E$ from AS61576 is \texttt{valid}, from AS61575 is \texttt{invalid}.
\end{enumerate}

A vantage point might initially choose the route to AS61576 for both prefixes $P_R$ and $P_E$. 
If the vantage point then switches to the route to AS61575 for prefix $P_E$ during
configuration C1, this indicates a preference for valid routes over invalid routes. 
An even stronger indicator of this policy is when the vantage point then again switches 
its route for $P_E$ to AS61576 when configuration C2 begins. 
We can attribute these route changes to a \texttt{prefer-valid} policy, rather than 
a \texttt{filter-invalid} policy, since we have ensured that vantage points
will choose invalid routes for $P_E$ from both origin AS.
This reasoning works the same way if the vantage point chooses the route to AS61575 for prefix $P_R$.

We can differentiate these policies by configuring announcements from \peering in such a way that a target network receives different combinations of \texttt{valid} and \texttt{invalid} prefixes through clients, peers, and providers, then observing its decisions.

There are subtleties in checking \texttt{prefer-valid} policies, as
a network is ``allowed'' to use ROA status as one part of checking how
preferred a path is, but, for example, it may prefer invalid peer routes
over valid provider routes but not over valid peer routes. We will
explore how best to capture these policies, building on our work on
using \peering to uncover (non-security-related) routing policies
\cite{anwar15interdomain}.

\subsection{Establishing a Monitoring Platform}

We will deploy our methods on a live, publicly available monitoring platform: \url{https://rov.rpki.net}.
Our measurements will run on an at least weekly basis, to report about the ongoing deployment of RPKI-based route origin validation and filtering.
We hope that these results will not only help researchers to better understand new security protocols but also operators to identify mistakes and consider motivation in deployment.

With a longitudinal rather than one-off study, we can evaluate the impact of efforts that, for example, routing registries, Internet exchange points, vendor updates, and operator organizations make to encourage BGP security adoption.
Measurements of the effect of such campaigns may yield a better understanding of how to spur uptake.
With an Internet-wide characterization, our data may inform best practices, encourage adoption, and reveal topics worthy of study, and provide the basis for understanding overall coverage and effectiveness.

\section{Conclusion}
\label{sec:conclusion}
In this paper, we discussed steps and results towards a rigorous methodology for measuring adoption of RPKI route validation and filtering.
We showed that BGP data sets that are incomplete with respect to peering relations---as are all available public data sets---challenge any method based on passive uncontrolled experiments.
We discussed several pitfalls.
We identified that traffic engineering, combined with negligent ROA configurations, are largely responsible for the routing differences between invalids and non-invalids.
To allow for more solid conclusions, we argue for controlled experiments.
In fact, our measurements in February, May, and August 2017 revealed three AS that already deploy RPKI-based filtering, which has been confirmed by the operators.

By controlling our own announcements, we can uncover policies proactively, yielding a richer understanding of adoption and configurations than is possible via passive observation of existing announcements; and potentially uncovering issues before they would otherwise manifest.
Our ongoing measurements focus on RPKI validation at locations with high impact such as Internet exchange points.
Notably AMS-IX started filtering invalid routes \emph{by default} at route servers on October~20,~2017.
Our public measurement platform will report on this.

\subsection*{Reproducibility}
\label{sec:conclusion_reproducibility}
We make all source code as well as data used for both our attempt at replication of the uncontrolled methodology of~\cite{gchss-trds-17} as well as our presented controlled methodology available at \url{https://github.com/RPKI/rov-measurement-code}.
Our platform to continuously monitoring ROV adoption will be publicly available via \url{http://rov.rpki.net}.

\subsection*{Acknowledgements}

We would like to thank the authors of \cite{gchss-trds-17} for answering our questions regarding their methodology and data set.
We gratefully acknowledge open discussions about RPKI filters with Job Snijders, Antonio Prado (AS59175), and operators of AS50300.
Thanks to CAIDA for organizing the BGP hackathon in 2016, which was
important to kick this collaboration off; and thanks to the RIPE NCC for
providing us temporarily Internet ressources during the beginning of this
journey.

We also thank the anonymous reviewers and Olivier Bonaventure for their constructive feedback.

This work was partially supported by a Google Faculty Research Award and the German Ministry of Education and Research (BMBF) within the project X-Check.

\balance
\bibliographystyle{ACM-Reference-Format}
\bibliography{rov-measurement-ccr}

\end{document}